\documentclass{llncs}
\usepackage{graphicx}  
\usepackage{subfigure} 
\usepackage{amsmath}   
\usepackage{amsfonts,fancybox}

\begin{document}
\title{Privacy-Preserving Social Network \\ 
      with Multigrained and Multilevel Access Control}
\author{YounSun Cho}
\institute{Purdue University \\
\email\{cho52\}@purdue.edu}
\maketitle
\begin{abstract}
I study two privacy-preserving social network graphs 
to disclose the types of relationships of connecting edges 
and provide flexible multigrained access control.
To create such graphs, my schemes employ the concept of secretaries 
and types of relationships.
It is significantly more efficient than 
those that using expensive cryptographic primitives.
I also show how these schemes can be used for multigrained access control
with various options.
In addition, I describe how much these schemes are resilient to 
infer the types of connecting edges. 
\end{abstract}

\section{Introduction}
Motivated by the fact that social networks (e.g., MySpace, Facebook, LinkedIn, Twitter and Buzz) 
have become extremely popular without being much considered its privacy issues. 
The main breaches of privacy in social networks are related with 
the relationships between people as well as the profile information of individuals.
In recent years, retrieving sensitive information associated with the data-analysis 
over social network graphs and its countermeasures have been studied 
\cite{NS09}, \cite{KNX08}, \cite{FP06}, \cite{LDG08}, \cite{BDK07},
\cite{HMJ07}, \cite{LT08}, \cite{ZP08}, \cite{ZG08}, \cite{LWL08}, \cite{FS09}. 
More seriously, one of the serious misuses of social networks is related with crimes.
In recent years, teenage girls were kidnapped by men who met in the Facebook \cite{KID1}, \cite{KID2}.

In a network point of view, nodes can be either individuals or other social entities,
and edges can be relationships between such nodes.
I can categorize privacy breaches into four groups: 
1) identity disclosure; 2) attribute disclosure; 3) link disclosure; and 4) content disclosure.
The identity disclosure occurs when the identity of an individual which corresponds to a node is revealed,
and the attribute disclosure happens when the attributes of an individual are revealed more accurately
than it would be possible before the data release. 
Attribute disclosure is usually inferred by identity disclosure.
The link disclosure occurs when the sensitive relationship or communication between two individuals are revealed
while the content disclosure happens when the sensitive data associated with each individual node is disclosed.

The best privacy-preserving system for social network over graphs should consider all these problems.
However, none of proposed systems could solve all these issues in one shot 
since protecting against each privacy breach may require different techniques.
Moreover, privacy-preserving system over the graphs and networks is very challenging in the following reasons.
Designing graph-modification algorithm to preserve privacy 
while providing the utility of the data is difficult 
since the nodes and the edges in a graph are all correlated each other. 
Thus, even single change of a node or an edge can affect the entire network.
Furthermore, quantifying the privacy is one of the biggest challenging research topics,
and the measurement for the privacy in a graph is much more challenging issue.
A graph itself includes rich information, but there is no standard method
to quantify the effect of the privacy gain or loss when a node or an edge
is changed.
Since two nodes in a graph are indistinguishable with respect to some structural metrics,
it is difficult to decide privacy models for graphs and the measurements of privacy breaches.
As a result, it is hard to model the background knowledge and the capability of attackers in this setting.
Besides, it is also challenging to model the behavior of participants involved in a social networks.
\subsection{My Results}
I am therefore interested in improving the privacy in social network and 
providing flexible multigrained access controls without using encryption or signature 
(which, of course, can be added as an optional security and privacy enhancement). 

Therefore, my proposed scheme is considerably faster than 
those based on traditional symmetric-key or public-key algorithms and 
can be applied to small devices with low storage and computational resources 
(e.g., smart phones or tablets).

In this report I describe two new approaches to improve the privacy for user's types of relationships
and provide multigrained access control to social network pages.
After a preliminary setup that involves creating nodes of a graph,
nodes a user simply connect to nodes of other users.
I analyze the privacy of my proposed schemes, 
and show the probability of inference for a type of an edge. 

\section{Related Work}
Liu \emph{et al.} \cite{LDG08} surveyed the recent researches on 
privacy-preserving data analysis over graphs and networks,
and their work is a good starting point to do research on this issue.
They categorized privacy breaches in a social network into three groups: 
1) identity disclosure; 2) link disclosure; and 3) content disclosure. 
More interesting thing in their work is that 
they also discussed the privacy issue occurring in multi-party distributed computing.
Backstrom et al. \cite{BDK07} defined two types of attacks, active and passive attackers, 
in an anonymized social network, 
and this model of attackers is widely adopted by other researchers.

\subsection{Identity Disclosure}
Since each individual is represented as a node in a graph, 
the identity of a node is disclosed in the network.
Thus a naive approach to hide the identity of a node is to 
replace the identity with a fake name such as pseudo-random name, and
this technique is called (naive) anonymization.

Since Sweeney and Samarati \cite{SS98} introduced the $k$-anonymity for data privacy in tables,
several variant methods such as $l$-diversity \cite{MKG06} and $t$-closeness \cite{LLV07} have been proposed.
However, I need to consider the structural information as well as data in a graph model
since the structure of a graph can be used to identify a target node 
(e.g., using a degree of a node). 
As a consequence, k-anonymity \cite{SS98} and its variants \cite{MKG06}, \cite{LLV07} 
are not appropriate for the preserving privacy in a graph model.

Hay \emph{et al.} \cite{HMJ07} defined $k$-candidate anonymity model for social network
such that for every structure query over the graph, there exist at least $k$ nodes that match the query. 
They then suggested a graph perturbation through modifying the graph by 
edge-deletions followed by edge-insertion. This approach could be an attacker's capability of re-identification
of identities; however, it could not support $k$-candidate anonymity.

Liu and Terzi \cite{LT08} suggested $k$-degree anonymous graph to reduce 
the possibility of an attacker's structural inference of a graph from the degree of each node.
A graph is k-degree anonymous graph, for every node there exists $k-1$ number of other nodes with the same number of degrees.
They argued that this was efficient anonymizing graph technique; however, 
creating such graph requires to probe all nodes and their edges respectively.
Moreover, the position of created nodes with same degrees should be considered.

Zou and Pei \cite{ZP08} described an algorithm for the subgraph created by the immediate $k$ neighbors of a node
based on greedy heuristics.
They defined $k$-neighborhood anonymity of a graph.
Given a graph $G$, for each node $u$, the algorithm finds $k-1$ neighbor nodes for $u$. 
For every pair of node $(u, v_i), i=1,\cdots,k-1$ and for each pair $(u, v_i)$,
the algorithm modifies the neighborhood subgraph of $u$ and the neighborhood subgraph of $v_i$
to make them isomorphic to each other. However, this algorithm is heuristic, moreover, expensive approach.
\subsection{Link Disclosure}
Zheleva and Getoor \cite{ZG08} addressed the problem of preserving privacy of sensitive relationship in a social network.
The authors introduced five different edge anonymization algorithms by adopting different removing edges
or clustering edges between nodes under the very specific assumptions.

Liu \emph{et al.} \cite{LWL08} considered edge weight in a graph and described two edge weight perturbation
algorithms for a different type of social network (e.g., transaction expenses could be the edge weight). 
They perturbed the edge weights before they disclose them in the graph.

Interesting observation is that most related work have been done here
with respect to analysis of disclose relationships in a social network \cite{NS09}, \cite{KNX08}, \cite{FP06}
instead of suggesting new countermeasures. 
Narayana and Shmatikov \cite{NS09} pointed out that the anonymization of a graph 
was not sufficient for privacy in a social network graph.
\subsection{Using Cryptographic Techniques}
Carminati \emph{et al.} \cite{CFP07} described an access control model 
where authorized users determined based on the relationships
are only granted to access to another user's social network.
They stored the encrypted relationship information with relationship
keys using public key cryptography.

Frikken and Srinivas \cite{FS09} suggested a key management scheme
in social networks without a trusted third party.
With the derived keys, it protects user's content as well as 
enhances access control to a social network.
A key for each user is derived based on the proximity of relationship between users.
Suppose Alice, a friend of Bob, can access Bob's social network with her friendship of Bob's key
and Carol, a friend of friend of Bob, can access Bob's social network 
with her friend of friend of Bob's key. 
Access can be controlled based on the those relationship keys 
that have different access policy of Bob's. 
This access control is done asynchronously, 
which means it does not require multiple users to be online simultaneously 
for the access control to allow access other users. 
For the social network graph protection, 
they encrypts destination nodes as well as edges in between.
One drawback as the authors mentioned is that
malicious users could access to the content of other unauthorized users 
by publishing other users' keys.

\section{Problem Statement and Goals}
Once Alice builds relationships with other users, she has connections with others.
Connections between users in social networks can be represented as a graph such that 
users can be represented as nodes and connections can be represented as edges.
These connections show that users have connected with other users, assuming as friends usually.
In current social networks, there are two types of users in terms of access grants: 
non-connected user (guest) and connected user(friend). 
Guests only can see limited profile or web pages based on Alice's privacy setting, 
while all connected users as friends have same access level to Alice's social web page.

In practice, however, Alice can have many types of relationships with others.
As an example, Alice actually consider Bob as an acquaintance, Carol as a competitor, 
and David as a business partner, Eve as an enemy, Flintz as a friend, and George as a family member, 
yet she does not want to disclose how she really thinks about others to keep good relationships 
without hurting other's feeling. Let us say this is social smiling.
This is sophisticated and delicate issue but critical information.

To achieve this goal, Alice hires secretaries and each secretary has two types of tags:
private and public name tags.
The former is for Alice's management purpose, and the latter is for connection with other users.
All secretaries of Alice have same public tags (e.g., friend or poker face), 
but a private tag of each secretary will be different based on an actual type of relationships, 
and only Alice can see and tell the differences.
When a connection established with another user, 
Alice assigns which type of secretary will actually handle the connection with the user.
By doing this, Alice can keep good relationships with others in public while she keeps her real thought in secret.
In addition, Alice can grant a different access grant to each connected user 
(e.g., by displaying different web pages based on the types of relationships).

\subsection{Adversary Model}
I assume that an adversary can access any public information (a graph)
and has a polynomial-time bounded computational power.
For threat models, I will consider the two types of adversaries: seeker and attacker.
A seeker is a person who explores web pages in a social network site and collects
(maybe sensitive) user information with (or without) registering the social networks.
This can be done easily since a default privacy setting in many cases allows everyone 
to read a user's profile (e.g., Facebook).

On the other hand, an attacker collects sensitive user information more aggressively and maliciously.
I consider two types of attackers on a social network \cite{BDK07}: active and passive attackers.
An active attacker creates many new connections with users (targets) on purpose 
and then tries to find some useful information with or without colluding with other active attackers.
However, a passive attacker does not have to create many new connections and just uses his already existing connections. 
He colludes with other passive attackers, exchanges some information learned from the existing connections,
and then tries to infer some sensitive information of target users. 

In practice, this will be useful to infer some small-scale sensitive information 
while the active attacker scenario might be appropriate to obtain useful information 
for a larger number of targeted users. 
Hence, for now I will more focus on the passive attacker for analyzing privacy in my scheme 
since passive attacker scenario is more realistic 
in a sense that any malicious one can achieve his small goal
without spending a large amount time and efforts.

\section{Proposed Scheme}
\subsection{Overview of Approach}
Suppose Alice considers Bob as an acquaintance, and Bob considers Alice as a competitor
while they disclose their actual thoughts each other.
First, when Alice and Bob register in a service provider (e.g., Facebook),
their secretaries are also created respectively. 
Alice generates $n$ number of secretaries $(s^1_a, s^2_a, \cdots, s^n_a)$.
And each secretary is in charge of different types of relationships such as friend, enemy, acquaintance, and etc.
For example, $s^a_1$ handles friends, $s^2_a$ handles enemies, $s^3_a$ handles acquaintances, 
$s^4_a$ handles friends, $s^5_a$ handles acquaintances, $\cdots$, $s^n_a$ handles friends.
Bob also creates $m$ number of secretaries $(s^1_b, s^2_b, \cdots, s^m_b)$.
$s^1_b$ handles enemies, $s^2_b$ handles friends, $s^3_b$ handles acquaintances, 
$s^4_b$ handles acquaintances, $s^5_b$ handles friends, $s^6_b$ handles competitors, $\cdots$, $s^m_b$ handles competitors.
Note that all users' initial number of secretaries could be either same or different,
and I set them differently, and thus $n \neq m$, 
and each secretary can handle multiple connections with other user's secretaries.

Second, since Alice considers Bob as an acquaintance, and Bob considers Alice as a competitor,
a connection between $s^3_a$ (or $s^5_a$) and $s^6_b$ (or $s^m_b$) will be established.
Thus the connections will be represented as $(Alice) - s^3_a - s^m_b - (Bob)$, 
denoted as $\{s_a^3({acquaintance}_a^1), s_b^{m}({competitor}_b^2)\}$ and so on.
The actual public connected edge is $s^3_a - s^m_b$ representing a friend.
Then Alice and Bob disclose this relationship graph to the public domain.
We can represent these relationships as a graph.

\subsection{Naive Graph Creation}
In this section, I will describe how a secretary-based privacy-preserving graph 
with multi-grained access control can be achieved.
Suppose a threshold value is $th_u$, and the number of types of relationships is $\tau_u$ for a node $u$.
\begin{itemize}
 \item Setup Phase: \\
  When a node $u$ is created\footnote{You can consider the creation of a node $u$ as 
	the registration of your email/password in the Facebook.}, 
  $n$ number of secretary nodes (snodes) $s^i_u$ where $i = 1 \cdots n_u$, and $\tau_u \leq n \leq th_u$ are created.
  For each type of relationship, $r_u(t)$,  $1 \cdots n_u / {\tau_u}$ snodes are assigned, 
  and thus a snode $s^i_u$ is responsible for a $r_u(t)$ type of relationship, denoted as $s^i_u \{ r_u(t) \}$.
  \item Connection Phase: \\
  Suppose a node $u$ wants to have a type $r_u(t)$ relationship with a node $v$, 
	and the node $v$ agrees with the connection to the node $u$ by choosing a type $r_v(t')$ relationship.
  Then a snode $s_u^i \{ r_u(t) \}$ will be connected as an edge with 
	 						 $s_v^{i'} \{ r_v(t') \}$ of a node $v$, 
  denoted as $[ s_u^i \{ r_u(t) \}, s_v^{i'} \{r_v(t') \} ]$.
	Note that $u$ does not know which type of relationship $v$ will choose and vice versa.
\end{itemize}
Notice that for each $u$ the number of secretaries can be different, and
it has a different type of relationship such as friend, enemy, acquaintance, and etc.

\subsection{Advanced Graph Creation}
\label{subsec:Advanced Graph Creation}
In the setup phase of the previous naive graph creation,
when a node $u$ is created, it simply assigns $n_u / \tau_u$ number of snodes to each type of relationship.
But this does not desirably reflect our real lives.
Suppose Alice has five types of relationships: friend, acquaintance, family, business, enemy.
Some types of relationships might need more number of secretaries than others
(e.g., in most cases, the number of friends are much larger than that of enemies).
Thus Alice wants to assign more number of secretaries for friend and acquaintance than the remaining types of relationships.
Moreover, assigning many secretaries to one particular type is not desirable to preserve privacy
since an adversary can infer that it could be friend or acquaintance 
when he sees connecting many edges to one particular node. 
Therefore, uniformly distribution of edges to each secretary 
will be more appropriate approach so as not to disclose types of relationship.

To achieve this requirement, let a node $u$ has $j$ numbers of $t$ type of relationship, 
and one of which has different $k(j)$ number secretary nodes, denoted as $r_u(t^j_{k(j)})$.
As an example, Alice has four friend relationships, three acquaintance relationships, 
one family relationship, two business relationships, and one enemy relationship.
Then she assigns 20 secretaries for the first friend relationship, and 10 secretaries for the second friend relationship, and
30 secretaries for the third and fourth friend relationship respectively and assigns 25 secretaries for enemy relationship and so on.
Note that for each node, a threshold value for the total number of snodes,
and the number of types of relationship could also be different 
since each node requires different levels of privacies.

I will use the following notations in this section.
\begin{itemize}
	\item $s_u^i$: A node $u$'s $i^{th}$ secretary node 
	\item $r_u(t^j_{k(j)})$: A node $u$ has $j$ numbers of $t$ type of relationship, 
				and one of which has different number $k(j)$ secretary nodes.
			  For example, suppose $r_u(t^j_{k(j)})$ has two friend relationship, 
				 $friend^1$ and $friend^2$, and $friend^1$ has 5 secretary nodes and $friend^2$ has 7 secretary nodes.
				 Then this is denoted as $r_u(friend^1_5)$ and $r_u(friend^2_7)$.
				I will just denote it as $r_u(t^j_{k})$ for convenience.
	\item $s_u^i \{ r_u(t^j_k) \}$: A secretary node $s_u^i$ is assigned to a relationship $r_u(t^j_k)$ 
\end{itemize}
Then the advanced graph will be created as follows.
\begin{itemize}
 \item Setup Phase: \\
  A snode $s_u^i$ is assigned to a relationship $r_u(t^j_k)$, denoted as $s_u^i \{ r_u(t^j_k) \}$
  shown in the notation. 
  \item Connection Phase: \\
  Suppose a node $u$ wants to have a type $t$ relationship $r(t)$ with a node $v$, 
	and a node $v$ wants to have a type $t'$ relationship $r(t')$ with a node $u$. 
  Then a snode $s_u^i\{ r(t^j_k) \}$ of the node $u$ will be connected with $s_v^{i'} \{ r(t'^{j'}_{k'}) \} $ of the node $v$, 
  denoted as $[s_u^i \{ r(t^j_k) \}, s_v^{i'} \{ r(t'^{j'}_{k'}) \} ]$.
\end{itemize}

\subsection{Comments and Notes}
I describe some comments and notes in this section.
\begin{itemize}
\item The more active user you are, the more secretaries you can hire 
			and more types of relationships you can create by employing the concept of incentive as necessary. 
			The definition of an active user could be dependent on a policy of each social network site. 
			For example, an active user could be defined by contributions to a service provider (e.g., Facebook)
			such as the number of logins, the time to use, the number of people involved in a user's social network,
			the number of posting advertisements or the number of clicking advertisements.
			Or this can be done, if a user is willing to pay additional cost, 
			then a service provider can provide more flexible and enhanced privacy as described above.
\item A user can distinguish his types of relationships of connecting edges with private tags,
			while the connecting edges disclose any information regarding types of relationships with one public tag
			of secretaries. The private tags of secretary nodes (snodes) are secret information to the public. 
			Thus an adversary does not know whether an edge represents acquaintance or enemy or something else.
\item	Each user can modify the types relationship of secretaries by swapping its role for other secretaries' role,
			and adding or deleting secretaries can be done easily since this can be done by a user's management of its directories or lists.
\end{itemize}
\subsection{Multigrained Access Control based on the Relationship}
Due to the fact that each user can manage his secretaries by creating them 
and respectively assigning, deassigning and reassigning their roles with private tags, 
each user can easily define the private access level to each type of relationships.
Alice's example in the previous sections, she can assign a different access grant to 
each group of people connecting to a particular type of relationships.
For a group of people who belong to her family, she allows them to see her personal pictures,
write some comments, and read her travel schedules.
For the people in the business group, she can let them read, write, download and post some professional articles,
and she can just display a bogus web page to a group of people in the enemy group. 
On the other hand, she can temporarily give an access to a simplified profile page 
without critical personal information such as address or phone number 
to a new member (e.g., a friend of friends of friends),
then after she has some trust to a new member, she can gradually give a higher grant to those members.
Notice that Alice also can define more sophisticated access controls by dividing a type of relationships into
sub-types of a relationship (e.g., friendship, for example, can be sub-divided into friend, close friend and best friend).

Another interesting possible scenario is as follows.
Alice has relationships with Bob and Carol, but she does not want them to know each other for some reason
through her social network, maybe because Bob and Carol are good business partners of Alice, 
but Alice does not lose a new business opportunity 
by letting them know each other and creating a new business without Alice.
Or Alice does not want to share valuable business opportunity between Bob and Carol. 
Currently, however, in social network sites (e.g., Facebook), 
as long as Bob and Carol have relationships with Alice, they can see each other in the friends' list of Alice,
and then they can also create a new relationship each other.

Our scheme can easily prevent such possibility by putting Bob and Carol in different types of relationships.
Although Bob and Carol are in a same type of relationships (e.g., friend), 
Alice can assign them to different friend groups, say friend1 and friend2, and
then can permit different grants to two friend groups so that they cannot see each other via Alice's social network web page.
She has another alternative to achieve the same goal 
by putting them in different sub-type of a relationship, say friend and best friends. 
As I explained in this section, my new scheme supports flexible multilevel and multi-grained access control 
without much difficulties. 

\section{Analysis of Privacy}
The scheme in this report provides privacy by employing the concept of secretaries, 
which hide the actual types relationship of a user from the public domain, 
yet allow a user to manage them efficiently and easily inside.
The more evenly the secretaries jobs distributed and bigger number of types of relationships, 
the more privacy the user can preserve with a denser and more evenly woven veil.
At the same time, however, secretaries are overheads in the view of management of a social network.
This overhead of managing secretaries for a user will be increased, 
but this is not significant and thus could be ignored.
However, the overhead for an entire social network cannot be ignored. 
Suppose each user generates $n$ secretaries and $k$ types of relationships, 
and $n/k$ secretaries are assigned to each type of relationship,
and each user has $c$ number social connections on average with other users.  
Let the total number of users in a social network be $m$.
Then $nm$ number of secretaries will be created in the entire social network.

As I mentioned, evenly distributed jobs to secretaries imply better privacy 
since denser connections to a few particular secretaries give a hint for a type of relationship to an adversary.
On the other hand, jobless secretaries result in an unnecessary redundancy.
To prevent this situation, in the naive graph creation scheme, 
each secretary is necessary to be in charge of $ck/n (=p)$ connections on average.  
Now I will consider the privacy level of a user for the naive graph creation scheme.
Given the number of secretary, $n$, 
the probability of guessing the number of type of relationships $k$ is $1/n$ since $1 \leq k \leq n$. 
Given the number of types of relationship, 
the probability of guessing a secretary's job is $1/k$.
Thus given a secretary node, the probability of guessing a type of relationships is $1/(kn)$
For the advanced graph creation scheme,
let $l$ be the number of same type relationship, 
and let each type of relationships has the same number of $l$ for the convenience of mathematical computation.
In the example of the notations in section \ref{subsec:Advanced Graph Creation}, 
there are two friend relationships, $friend^1$ and $friend^2$, and thus $l$ is 2 in this case.
We can consider $k l$ as an the number of types of relationships, denoted as $k' = k l$.
Then given a secretary node, the probability of guessing a type of relationships is $1/(k l n)$

Based on this analysis, the level of privacy is more affected by the number of types of relationships 
rather than the number of secretaries. 
Thus a user's real thought is veiled by the increased number of types of relationships 
with an optimized number of secretaries such that $k \leq n$.
Given $k$ types of relationships, 
Alice can enhance her privacy level by increasing the number of same type relationship, $l$,
and creating the sub-types of each relationship,
resulting in the number of lists or directories of which Alice has to manage
while minimizing the burden of an entire social network.

\section{Future Work and Conclusion}
I have described two schemes - naive graph creation and advanced graph creation - 
to improve the privacy of social network and provide multilevel and multigrained access control
by adopting the concept of secretaries.
The overhead of adding and managing secretaries for each user might be ignored, 
however, the overhead for an entire social network needs to be minimized.
Both schemes provide privacy for types of connecting relationships 
with low storage and computational power for a small device.
I also analyzed an optimized total number of secretaries in an entire social network.

In terms of further exploration of the vulnerability of privacy to the active attack, 
one possible direction for future work would be to use inexpensive cryptographic primitives 
(e.g., hash function or symmetric key cryptography).
In addition, other directions for future research could exploit different privacy measurements
for my schemes. 
\bibliographystyle{plain}
\bibliography{ppsn-mmac}

\begin{thebibliography}{10}

\bibitem{KID2}
Facebook kidnapped in uk.
\newblock Available at
  http://www.guardian.co.uk/uk/2010/mar/08/peter-chapman-facebook-ashleigh-hal%
l.

\bibitem{KID1}
Facebook kidnapped in us.
\newblock Available at
  http://blog.taragana.com/index.php/archive/california-man-accused-of-kidnapp%
ing-15-year-old-idaho-girl-he-met-on-social-networking-site/.

\bibitem{BDK07}
L.~Backstrom, C.~Dwork, and J.~M. Kleinberg.
\newblock Wherefore art thou r3579x?: Anonymized social networks, hidden
  patterns, and structural steganography.
\newblock In {\em In Proceedings of the 16th International Conference on World
  Wide Web (WWW’07)}, pages 181--190, Alberta, Canada, May 2007.

\bibitem{CFP07}
B.~Carminati, E.~Ferrari, and A.~Perego.
\newblock Private relationships in social networks.
\newblock In {\em In Private Data Management Workshop (held in conjunction with
  ICDE’07)}, Istanbul, Turkey, April 2007.

\bibitem{FS09}
K.~Frikken and P.~Srinivas.
\newblock Key allocation schemes for private social networks.
\newblock In {\em Workshop on Privacy in the Electronic Society (WPES'09)},
  2009.

\bibitem{FP06}
K.~B. Frikken and P.~Golle.
\newblock Private social network analysis: how to assemble pieces of a graph
  privately.
\newblock In {\em WPES}, page 89–98. ACM, 2006.

\bibitem{HMJ07}
M.~Hay, G.~Miklau, D.~Jensen, P.~Weis, and S.~Srivastava.
\newblock Tapestry: An infrastructure for fault-tolerant wide-area location and
  routing.
\newblock Technical Report Technical Report No. 07-19, University of
  Massachusetts Amherst, Computer Science Department, March 2007.

\bibitem{KNX08}
A.~Korolova, S.~Nabar, Y.~Xu, and R.~Motwani.
\newblock Link privacy in social networks.
\newblock In {\em Proceedings of the 21st International Conference on Data
  Engineering (ICDE)}, 2008.

\bibitem{LLV07}
N.~Li, T.~Li, and S.~Venkatasubramaniany.
\newblock t-closeness: Privacy beyond k-anonymity and l-diversity.
\newblock In {\em In ICDE ’07: Proceedings of the 23rd International
  Conference on Data Engineering}, page 106–115, Washington, DC, USA, 2007.

\bibitem{LDG08}
K.~Liu, K.~Das, T.~Grandison, and H.~Kargupta.
\newblock {\em Privacy-Preserving Data Analysis on Graphs and Social Networks}.
\newblock Chapman \& Hall/CRC, New York, NY, December 2008.

\bibitem{LT08}
K.~Liu and E.~Terzi.
\newblock Towards identity anonymization on graphs.
\newblock In {\em In ACM SIGMOD International Conference on Management of Data
  (SIGMOD'08)}, pages 93--106, Vancouver, Canada, June 2008.

\bibitem{LWL08}
L.~Liu, J.~Wang, J.~Liu, and J.~Zhang.
\newblock Privacy preserving in social networks against sensitive edge
  disclosure.
\newblock Technical Report Technical Report CMIDA-HiPSCCS 006-08, Department of
  Computer Science, University of Kentucky, 2008.

\bibitem{MKG06}
A.~Machanavajjhala, D.~Kifer, J.~Gehrke, and M.~Venkitasubramaniam.
\newblock L-diversity: Privacy beyond k-anonymity.
\newblock In {\em ACM Trans. Knowl. Discov. Data, 1(1):3}, 2006.

\bibitem{NS09}
A.~Narayanan and V.~Shmatikov.
\newblock De-anonymizing social networks.
\newblock In {\em S\&P 2009}, 2009.

\bibitem{SS98}
P.~Samarati and L.~Sweeney.
\newblock Generalizing data to provide anonymity when disclosing information.
\newblock In {\em In Proc. of the 17th ACM-SIGMOD-SIGACT-SIGART Symposium on
  the Principles of Database Systems}, 1998.

\bibitem{ZG08}
E.~Zheleva and L.~Getoor.
\newblock Preserving the privacy of sensitive relationships in graph data.
\newblock In {\em Proceedings of the First International Workshop on Privacy,
  Security, and Trust in KDD}, pages 153--171, 2007.

\bibitem{ZP08}
B.~Zhou and J.~Pei.
\newblock Preserving privacy in social networks against neighborhood attacks.
\newblock In {\em 2008 IEEE 24th International Conference on Data Engineering
  (IEEE, 2008)}, pages 506--515, 2008.

\end{thebibliography}

\end{document}